\documentclass[conference]{IEEEtran}
\IEEEoverridecommandlockouts

\usepackage[utf8]{inputenc}
\usepackage[T1]{fontenc}
\usepackage{tcolorbox}
\usepackage{xcolor}
\usepackage{graphicx}
\usepackage{booktabs}
\usepackage{multirow}
\usepackage{array}
\usepackage{tabularx}

\usepackage{enumitem}
\usepackage{amsmath,amssymb}
\usepackage{bm}
\usepackage{algorithm}
\usepackage{algpseudocode}
\usepackage{tikz}
\usepackage[hyphens]{url}

\usepackage{xurl}                    
\usepackage[hidelinks]{hyperref}     
\hypersetup{breaklinks=true}

\def\BibTeX{{\rm B\kern-.05em{\sc i\kern-.025em b}\kern-.08em
    T\kern-.1667em\lower.7ex\hbox{E}\kern-.125emX}}

\title{Towards a Humanized Social-Media Ecosystem:\\
AI-Augmented HCI Design Patterns for Safety, Agency \& Well-being}

\author{
\IEEEauthorblockN{1\textsuperscript{st} Mohd Ruhul Ameen}
\IEEEauthorblockA{\textit{College of Engineering and Computer Sciences}\\
\textit{Marshall University}\\
Huntington, WV, USA\\
ameen@marshall.edu}
\and
\IEEEauthorblockN{2\textsuperscript{nd} Akif Islam}
\IEEEauthorblockA{\textit{Department of Computer Science and Engineering}\\
\textit{University of Rajshahi}\\
Rajshahi 6205, Bangladesh\\
iamakifislam@gmail.com}
}

\begin{document}
\maketitle

\begin{abstract}
Social platforms connect billions of people, yet their engagement-first algorithms often work on users rather than with them, amplifying stress, misinformation, and a loss of control. We propose Human-Layer AI (HL-AI) which is a user-owned, explainable intermediaries that sit in the browser between platform logic and the interface. HL-AI gives people practical, moment-to-moment control without needing platform cooperation. We contribute a working Chrome/Edge prototype that implements five representative patterns framework (Context-Aware Post Rewriter, Post Integrity Meter, Granular Feed Curator, Micro-Withdrawal Agent and Recovery Mode), a unifying mathematical formulation that balances user utility, autonomy costs and risk thresholds, and an evaluation plan spanning technical accuracy, usability and behavioral outcomes. The result is a suite of humane controls that help users rewrite before harm, read with integrity cues, tune feeds with intention, pause compulsive loops and seek shelter during harassment while preserving agency through explanations and override options. We argue that HL-AI offers a practical path to retrofit today’s feeds with safety, agency and well-being, and we release design patterns and implementation details to support adoption and future study. As a prototype, this work invites rigorous user evaluation across cultures and contexts.
\end{abstract}


\section{Introduction}

Social media platforms have fundamentally reshaped human communication and information access, with more than 5 billion active user identities worldwide as of 2024~\cite{datareportal2024global}. While these platforms have enabled unprecedented connectivity, knowledge sharing, and civic participation, their underlying architectural paradigm remains rooted in \emph{simplex communication}---a unidirectional system in which platform-controlled algorithms curate and prioritize information flows without meaningful user input~\cite{metzler2024social, sun2024right}. This engagement-optimized model inherently prioritizes attention capture over human-centered outcomes such as psychological wellbeing, information integrity, and meaningful task accomplishment.

The societal consequences of this misalignment have become increasingly evident across South Asia. The July 2025 Milestone School helicopter crash in Bangladesh demonstrated how algorithmic feeds can exacerbate psychological trauma by repeatedly surfacing distressing content in the absence of user-controlled filtering mechanisms~\cite{dhakatribune2025milestone, cnn2025milestone}. Survivors and community members were continually exposed to graphic imagery and updates, intensifying mental health distress as feeds optimized for engagement rather than emotional protection~\cite{dailystar2025milestone}.

Similarly, during Bangladesh's July Revolution of 2024---when student-led protests culminated in the resignation and flight of Prime Minister Sheikh Hasina on August 5, 2024---social media platforms became simultaneous tools for democratic mobilization and vectors for large-scale misinformation~\cite{aljazeera2024hasina, wikipedeia2024julyrevolution}. Platform-driven amplification dynamics contributed to volatility during the uprising, in which over 1,400 people were killed~\cite{usip2024bangladesh}.

More recently, Nepal's September 2025 Gen Z protests highlighted both the potential and structural limitations of current social platforms. Following the government's ban of 26 major social media services, protesters migrated to Discord, where over 145,000 participants used structured discussions and polls to democratically select an interim Prime Minister~\cite{aljazeera2025nepal, wikipedia2025nepal, npr2025nepal}. This event demonstrated a clear demand for \emph{duplex communication}---systems that empower users to actively shape their information ecosystem rather than passively receiving algorithmically curated feeds.

Compounding these issues, cross-border misinformation campaigns have intensified regional tensions. In 2024, Indian media outlets circulated more than 137 false reports about Bangladesh, many algorithmically amplified on social platforms~\cite{rumorscanner2025indian, voa2024tensions}. These incidents illustrate how engagement-driven architectures can propagate harmful narratives across national boundaries, producing real-world diplomatic and communal consequences.

Despite growing awareness of these systemic issues, current mitigation approaches remain fundamentally constrained by the simplex communication paradigm. Existing mechanisms---such as basic screen-time limits, opaque feed adjustments, or platform-controlled moderation systems---do not grant users meaningful control over how information is curated, prioritized, or de-emphasized. Users increasingly expect granular, value-aligned control over their digital environments: for example, reducing political content while increasing educational material, or limiting exposure to sensitive topics during vulnerable periods~\cite{narayanan2023understanding}.

At present, however, mainstream platforms lack architectural support for \textit{user-owned, explainable AI intermediaries} that can mediate between raw platform outputs and the user interface. This architectural gap prevents the development of systems that support real-time transparency, personalization, and psychological wellbeing.

The consequences are measurable and widespread. Studies report that:
\begin{itemize}
\item 62\% of daily users experience post-session regret,
\item trust in platform-mediated content has declined to 28\%, and
\item 38.2\% of users accidentally share misinformation due to inadequate decision-support mechanisms~\cite{watson2023social}.
\end{itemize}

Events such as the Milestone School tragedy~\cite{dhakatribune2025milestone} and Bangladesh–India misinformation campaigns~\cite{bbc2024fake} highlight how the absence of user-controlled intermediaries can exacerbate trauma, distort information ecosystems, and intensify geopolitical tensions.

Research in human-centered AI has established foundational principles for promoting user autonomy, transparency, and controllability~\cite{garibay2023six, stephanidis2023hci}. Parallel work in digital wellbeing identifies the psychological and behavioral factors that shape healthy technology use and proposes various intervention mechanisms~\cite{buchi2024digital, internetmatters2024children}. Advances in explainable AI (XAI) have further introduced techniques for interpreting model behavior and making algorithmic decisions more intelligible to users~\cite{scharowski2023exploring, liao2022human}. Despite these developments, existing approaches remain constrained by several critical limitations. First, digital wellbeing tools predominantly rely on restriction-based mechanisms—such as time limits or app blocking—rather than augmentation-based strategies that enhance users’ ability to make informed decisions~\cite{molnar2024explainable}. Second, misinformation detection and content ranking infrastructures remain centrally controlled by platforms, offering limited transparency and virtually no customization for end users~\cite{goldstein2023generative, pilati2025artificial}. Third, most explainable AI methods provide post-hoc explanations that describe algorithmic decisions only after they occur, rather than enabling proactive, user-controlled intermediation capable of shaping information flows in real time~\cite{vilone2024comprehensive}. Collectively, these shortcomings reveal a fundamental research gap: the absence of comprehensive design patterns and architectural frameworks for developing \emph{user-controlled AI intermediaries} that can operate within social media environments.

This paper contributes a practical and human-centered approach to improving user control and wellbeing in social media systems. First, we introduce an architectural model for user-owned AI intermediaries that operate between platform algorithms and the user interface, enabling greater transparency and autonomy. Second, we present a taxonomy of fifteen Human-Layer AI (HL-AI) design patterns spanning safety, agency, and wellbeing; while the complete set will appear in a forthcoming journal publication, this paper illustrates five representative patterns to convey the core concepts. Third, we outline an initial mathematical formulation for balancing user autonomy and risk mitigation, providing a structured basis for customizing intermediary behavior. Finally, although full empirical evaluation is reserved for future work, we describe our planned methodology and demonstrate a working prototype that implements the five showcased design patterns.

\section{Related Work}

Human-Centered AI emphasizes transparency, explainability, and user agency~\cite{stephanidis2023hci}. Gausen et al.~\cite{gausen2025sociotechnical} introduced "sociotechnical transparency" considering both technical systems and their interactions with users. Recent work shows transparency significantly mitigates negative relationships between AI attitudes and trust when user involvement is high~\cite{ai_transparency_trust_2025}.

Digital Wellbeing Research evolved from restriction-based to augmentation-based frameworks enhancing user decision-making~\cite{monge2023digital}. Vanden Abeele's dynamic systems approach recognizes digital wellbeing as contingent on personal characteristics and design choices~\cite{vanden2020digital}. The METUX framework identifies autonomy, competence, and relatedness as essential for sustained motivation and wellbeing~\cite{peters2018metux}.

AI-powered moderation reports 99\% terrorism-related post detection before user reports~\cite{eu_ai_content_moderation_2025}. However, smaller accounts disproportionately spread AI-generated misinformation~\cite{proellochs2025characterizing}. Analysis reveals both downstream (post-spread identification) and upstream (prevention) approaches are needed.

Legal scholarship establishes "right to know" algorithms as fundamental to democratic participation~\cite{sun2024right}. Research shows granular control mechanisms beyond simple on/off switches are needed~\cite{ai_transparency_trust_2025}. Recent workshops identified novel design challenges for human-AI collaboration.

\section{Core Design Patterns}

\begin{figure}[!b]
    \centering
    \includegraphics[width=0.8\linewidth]{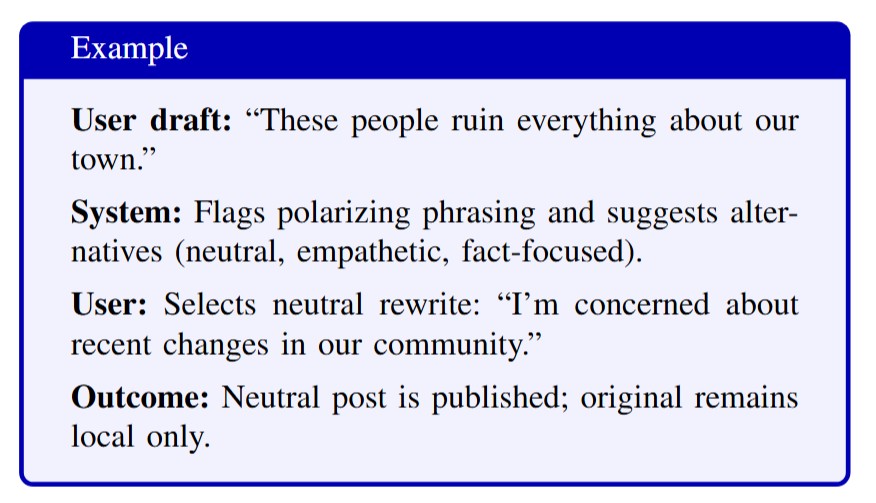}
    \caption{Example of P1. Context-Aware Post Rewriter}
    \label{fig:example_p1}
\end{figure}

\begin{table}[!t]
\caption{P1 — Context-Aware Post Rewriter}
\label{tab:p1}
\renewcommand{\arraystretch}{1.2}
\setlength{\tabcolsep}{3pt}
\centering
\begin{tabular}{p{0.22\linewidth} p{0.72\linewidth}}
\toprule
\textbf{Intent} &
Encourage reflection and reduce harmful posting by providing AI-assisted rewrites. \\
\midrule

\textbf{Problem} &
Users may unintentionally post polarizing or biased phrasing that escalates conflict. \\
\midrule

\textbf{Context} &
Fast-paced posting environments, public visibility, and sensitive sociocultural topics. \\
\midrule

\textbf{Forces} &
Reflection vs.\ speed; safety vs.\ autonomy; privacy vs.\ proactive guidance. \\
\midrule

\textbf{Solution} &
The intermediary reads the draft, generates neutral or empathetic alternatives, provides contextual warnings, and allows users to override suggestions at any time. \\
\midrule

\textbf{Consequences} &
\textbf{Pros:} Reduced harm and more mindful posting.  
\newline
\textbf{Risks:} May trigger perceptions of censorship or lead to over-reliance on AI rewrites. \\
\bottomrule
\end{tabular}
\end{table}

We present five representative patterns from our broader fifteen-pattern framework. These illustrate how Human-Layer AI (HL-AI) intermediaries enhance user safety, agency, and wellbeing.

\subsection{P1. Context-Aware Post Rewriter}
\label{subsec:p1}

This pattern aims to support reflective posting by offering AI-generated rewrites for potentially harmful content. The intermediary analyzes the draft, previews how it may be interpreted, and provides neutral or empathetic alternatives while leaving final choice to the user.

\subsection{P2. Post Integrity Meter}
\label{subsec:p2}

This pattern provides real-time integrity signals for posts by checking factual consistency, AI-generation likelihood, and political bias. Instead of restricting access, the intermediary offers transparent cues that help users judge credibility.

\begin{figure}[!t]
    \centering
    \includegraphics[width=0.8\linewidth]{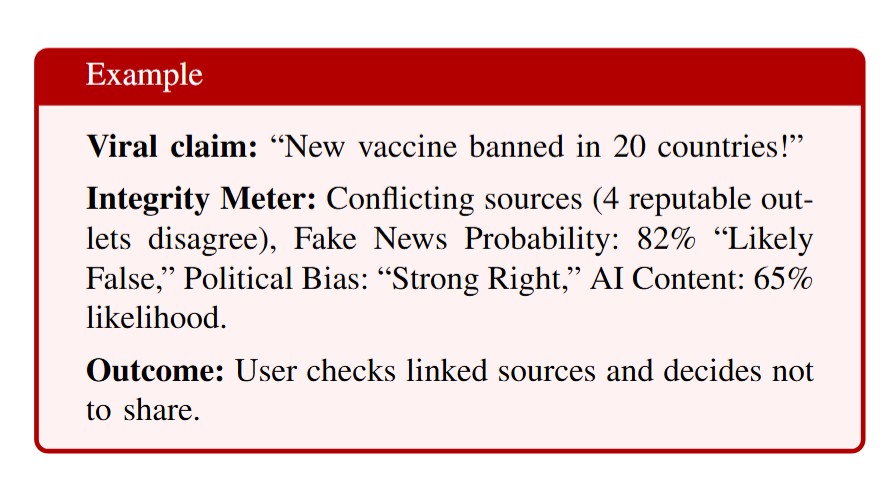}
    \caption{Example of P2. Post Integrity Meter}
    \label{fig:example_p2}
\end{figure}

\begin{table}[!t]
\caption{P2 — Integrity Meter}
\label{tab:p2}
\renewcommand{\arraystretch}{1.2}
\setlength{\tabcolsep}{3pt}
\centering
\begin{tabular}{p{0.22\linewidth} p{0.72\linewidth}}
\toprule

\textbf{Intent} &
Provide clear, real-time integrity cues showing whether content may be fake, AI-generated, or biased, supporting informed decision-making. \\
\midrule

\textbf{Problem} &
Users struggle to distinguish reliable posts from misinformation or synthetic media in fast-moving environments. \\
\midrule

\textbf{Context} &
High-stakes domains such as health, elections, and crises; mixed content streams; rapid skimming with minimal fact-checking. \\
\midrule

\textbf{Forces} &
Transparency vs.\ overload; automation vs.\ accuracy; awareness vs.\ trust; fast processing vs.\ verifiability. \\
\midrule

\textbf{Solution} &
The system introduces an Integrity Meter beneath each post. It checks claim consistency against reputable sources, estimates misinformation probability, detects AI-generated text, images, or videos, and provides contextualized bias estimations. All results are delivered through plain-language explanations with adjustable sensitivity to suit varying user preferences. \\
\midrule

\textbf{Consequences} &
\textbf{Pros:} Supports critical literacy and reduces misinformation spread.  
\newline
\textbf{Risks:} Users may over-trust probability scores, misinterpret bias labels, or face adversarial adaptation. \\
\bottomrule

\end{tabular}
\end{table}

\subsection{P3. Granular Feed Curator}
\label{subsec:p3}

This pattern gives users actionable control over feed composition and ad categories through simple sliders and toggles. Rather than relying on opaque algorithmic defaults, users can directly shape what they see and how often.

\begin{figure}[!b]
    \centering
    \includegraphics[width=0.8\linewidth]{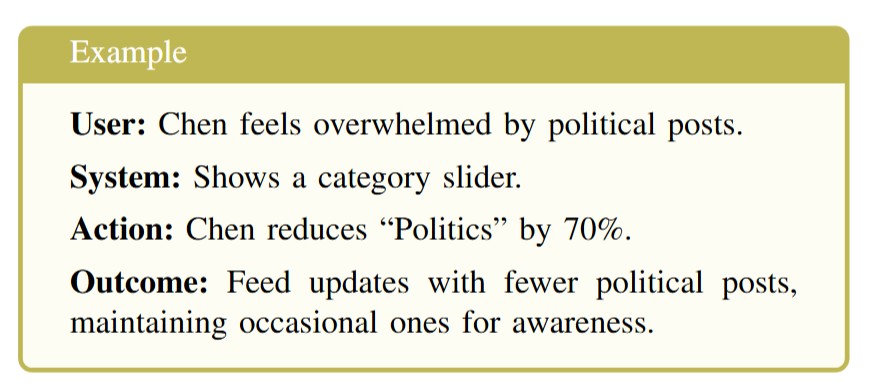}
    \caption{Example of P3. Granular Feed Curator}
    \label{fig:example_p3}
\end{figure}

\begin{table}[!b]
\caption{P3 — Granular Feed Curator}
\label{tab:p3}
\renewcommand{\arraystretch}{1.2}
\setlength{\tabcolsep}{3pt}
\centering
\begin{tabular}{p{0.22\linewidth} p{0.72\linewidth}}
\toprule

\textbf{Intent} &
Provide intuitive, fine-grained controls for adjusting feed content and ad exposure. \\
\midrule

\textbf{Problem} &
Users feel feeds are manipulated or unclear. Built-in platform controls are vague, buried, or ineffective, reducing trust and limiting personalization. \\
\midrule

\textbf{Context} &
Social feeds, recommender systems, and ad personalization scenarios where users want control without losing discovery. \\
\midrule

\textbf{Forces} &
Transparency vs.\ complexity; discovery vs.\ control; personalization vs.\ privacy; stability vs.\ adaptability. \\
\midrule

\textbf{Solution} &
The system provides a granular feed-filter module that lets users adjust content intensity across categories such as politics, sports, or personal updates. It offers an ad-transparency panel with options to disable specific ad categories, quick contextual toggles (e.g., focusing on friends), and per-post overrides that shape future recommendations. Auto-tuning powered by an HL-AI Curator Agent refines personalization while maintaining user agency. \\
\midrule

\textbf{Consequences} &
\textbf{Pros:} Higher trust, better relevance, reduced frustration.  
\newline
\textbf{Risks:} Over-curation may create echo chambers; extensive controls may overwhelm some users. \\
\bottomrule

\end{tabular}
\end{table}

\subsection{P4. Micro-Withdrawal Agent}
\label{subsec:p4}

This pattern introduces brief, context-aware pauses that help users step out of compulsive scrolling loops without imposing hard blocks. The intermediary detects continuation risk and offers lightweight reflective prompts or redirections.

\begin{figure}
    \centering
    \includegraphics[width=0.8\linewidth]{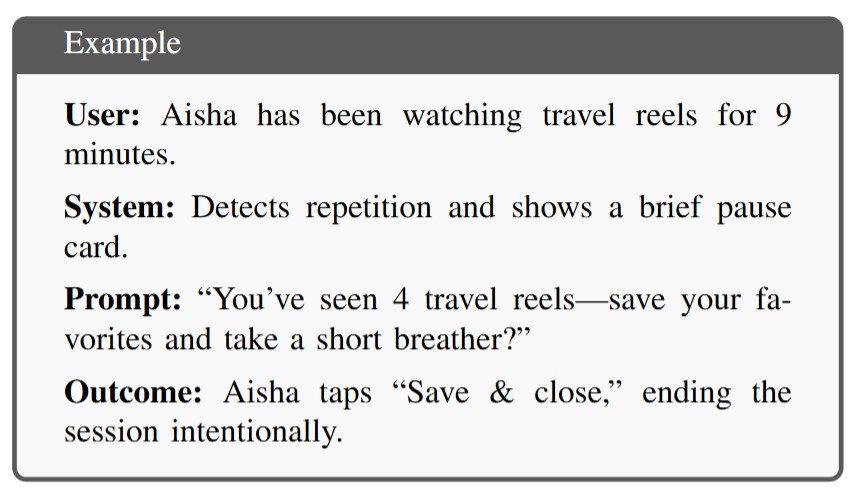}
    \caption{Example of P4. Micro-Withdrawal Agent}
    \label{example_p4}
\end{figure}

\begin{table}[!b]
\caption{P4 — Micro-Withdrawal Agent}
\label{tab:p4}
\renewcommand{\arraystretch}{1.2}
\setlength{\tabcolsep}{3pt}
\centering
\begin{tabular}{p{0.22\linewidth} p{0.72\linewidth}}
\toprule

\textbf{Intent} &
Provide brief, context-aware pauses that help users disengage from compulsive scrolling without paternalistic blocking. \\
\midrule

\textbf{Problem} &
Variable-ratio reward loops (“just one more”) drive overuse, regret, and fatigue. Traditional screen-time warnings are blunt and ignore situational factors. \\
\midrule

\textbf{Context} &
Infinite-scroll feeds, late-night usage, repetitive content cycles, and sessions that drift away from user goals such as staying informed, connecting with friends, or learning. \\
\midrule

\textbf{Forces} &
Relief vs.\ annoyance; agency vs.\ protection; behavioral inference vs.\ privacy; minimal disruption vs.\ effectiveness. \\
\midrule

\textbf{Solution} &
The system detects continuation risk by analyzing patterns such as swipe velocity, repetitive topic exposure, time-of-day signals, and stated user goals. When the likelihood of compulsive continuation is high, it delivers a brief micro-intervention lasting under ten seconds—providing a moment of reflection, an optional redirection to saved or meaningful content, or the frictionless option to continue. The cadence of these interventions adapts dynamically based on user responses, including acceptance, dismissal, or repeated avoidance. \\
\midrule

\textbf{Consequences} &
\textbf{Pros:} Reduces impulsive loops and increases goal alignment.  
\newline
\textbf{Risks:} Habituation, banner blindness, or perceptions of nagging. \\
\bottomrule

\end{tabular}
\end{table}

\subsection{P5. Recovery Mode}
\label{subsec:p5}

This pattern provides an immediate “shelter-in-feed” mode when users face harassment, sudden pile-ons, or emotional overload. The intermediary reduces harmful exposure, limits inbound interaction, and offers structured next steps.

\begin{figure}[!t]
    \centering
    \includegraphics[width=0.8\linewidth]{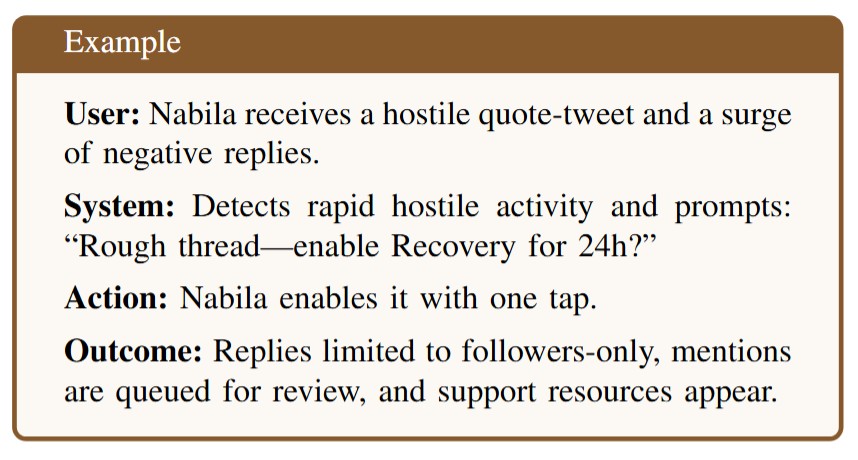}
    \caption{Example of P5. Recovery Mode}
    \label{fig:example_p5}
\end{figure}

\begin{table}[!t]
\caption{P5 — Recovery Mode}
\label{tab:p5}
\renewcommand{\arraystretch}{1.2}
\setlength{\tabcolsep}{3pt}
\centering
\begin{tabular}{p{0.22\linewidth} p{0.72\linewidth}}
\toprule
\textbf{Intent} &
Provide an immediate, protective mode after harassment or overload that reduces exposure, limits inbound interactions, and surfaces appropriate support. \\
\midrule

\textbf{Problem} &
Following negative events, users face toxic replies, brigading, and decision paralysis. Default feeds continue amplifying harm instead of stabilizing the situation. \\
\midrule

\textbf{Context} &
Online pile-ons, personal attacks, doxxing attempts, viral controversies, and vulnerable timing (late-night, travel). \\
\midrule

\textbf{Forces} &
Rapid safety vs.\ missing supportive messages; automation vs.\ consent; visibility vs.\ protection; shielding vs.\ information flow. \\
\midrule

\textbf{Solution} &
Recovery Mode enables one-tap activation—or an optional proactive suggestion—that immediately restricts who can reply or send direct messages, suppresses mentions, and hides toxic responses. It queues reports with automatically captured evidence, opens a dedicated support hub with trusted contacts and relevant guidance, and assists the user in planning a temporary exit such as short mutes or a timed account pause. \\
\midrule

\textbf{Consequences} & 
\textbf{Pros:} Rapid harm reduction; clearer decisions; preserved evidence.  
\newline
\textbf{Risks:} May filter out supportive messages; false positives could cause unnecessary isolation. \\

\bottomrule
\end{tabular}
\end{table}

\section{Mathematical Framework}

We formalize the trade-off between user autonomy and risk mitigation as an optimization problem. For each action or intervention $a$ associated with a content item, the system selects the action that maximizes the objective

\begin{equation}
\mathcal{J}(a) = u(a) - \lambda\,\Omega(a) - \beta\,\mathbf{1}_{\,r(a) > \tau}\,r(a),
\end{equation}

where $u(a)$ denotes the user-perceived utility of the action, $\Omega(a)$ represents the degree to which the intervention may compromise user agency, and $r(a)$ reflects the estimated risk associated with the content or behavior. The parameters $\lambda$ and $\beta$ weight the importance of preserving autonomy and mitigating risk, respectively, while $\tau$ represents the user-adjustable tolerance threshold above which risk incurs a penalty. This formulation encourages the system to favor actions that provide meaningful benefit while avoiding excessive interference unless a user's defined risk threshold has been exceeded.

Each of the five patterns instantiates this model through three common mechanisms. First, a transparency layer ensures that every intervention is accompanied by a short explanation and the option to override, preserving user control even when risks are flagged. Second, the model allows users to calibrate the autonomy–risk balance by adjusting parameters such as $\lambda$ and $\tau$, enabling different levels of sensitivity depending on personal goals, context, and comfort. Finally, an audit trail captures interventions and user responses, supporting accountability and enabling the system to adapt future behavior in a predictable and user-aligned manner. Together, these elements provide a principled mathematical basis for HL-AI intermediaries that align safety, autonomy, and wellbeing.

\section{System Architecture and Algorithms}

\subsection{HL-AI Master Decision Flow}

\begin{figure}[!t]
    \centering
    \includegraphics[width=0.49\linewidth]{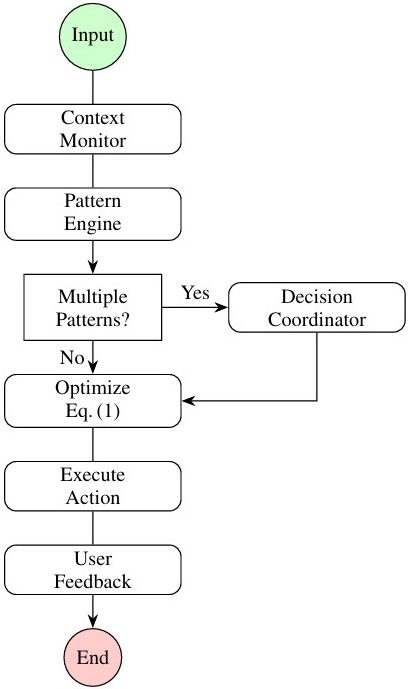}
    \caption{HL-AI master decision flow: pattern coordination and optimization.}
\label{fig:master_flow}
\end{figure}

Figure~\ref{fig:master_flow} illustrates the core decision-making process for the Human-Layer AI system, showing how multiple patterns coordinate through the centralized optimization framework.


\subsection{Core Optimization Algorithm}

Algorithm~\ref{alg:optimization} implements the mathematical framework from Equation (1), balancing user agency with risk reduction across all patterns.

\begin{algorithm}[!b]
\caption{HL-AI Pattern Optimization Framework}
\label{alg:optimization}
\footnotesize
\begin{algorithmic}[1]
\Require User preference weight $\lambda$, risk threshold $\tau$, content/action set $\mathcal{A}$
\Ensure Optimal intervention decision $a^{*}$

\State Initialize utility estimator $\hat{u}(a)$ and risk estimator $\hat{r}(a)$
\State Compute agency penalty $\Omega(a)$ according to intervention type

\For{each candidate action $a_i \in \mathcal{A}$}
    \State $J(a_i) \gets \lambda \cdot \hat{u}(a_i) + (1 - \lambda) \cdot \hat{r}(a_i) - \Omega(a_i)$
    \If{$\hat{r}(a_i) > \tau$ \textbf{and} intervention\_required}
        \State $J(a_i) \gets J(a_i) - \beta \cdot \hat{r}(a_i)$ \Comment{Apply safety penalty}
    \EndIf
\EndFor

\State $a^{*} \gets \arg\max_{a_i} J(a_i)$
\State \Return $a^{*}$ along with explanation and user override option
\end{algorithmic}
\end{algorithm}

\subsection{Post Integrity Assessment Algorithm}

Algorithm~\ref{alg:integrity} details the multi-signal approach for Pattern P3 (Post Integrity Meter), combining fact-checking, AI detection, and bias estimation.

\begin{algorithm}[h]
\caption{Multi-Signal Post Integrity Assessment}
\label{alg:integrity}
\footnotesize
\begin{algorithmic}[1]
\Require Post content $p$, fact-check database $\mathcal{D}$, AI detection model $M_{AI}$
\Ensure Integrity score vector $\mathbf{s} = [s_{fact}, s_{AI}, s_{bias}]$
\State Extract claims $C = \{c_1, c_2, ..., c_n\}$ from $p$ using NER
\State $conflicts \leftarrow 0$, $total\_claims \leftarrow |C|$
\For{each claim $c_i \in C$}
    \State $sources \leftarrow$ query\_database($c_i$, $\mathcal{D}$)
    \If{$|sources| > 0$ \textbf{and} any source contradicts $c_i$}
        \State $conflicts \leftarrow conflicts + 1$
    \EndIf
\EndFor
\State $s_{fact} \leftarrow 1 - \frac{conflicts}{max(total\_claims, 1)}$ \Comment{Fact score}
\State $s_{AI} \leftarrow M_{AI}(p)$ \Comment{AI-generated probability}
\State $s_{bias} \leftarrow$ estimate\_political\_lean($p$) \Comment{Bias classification}
\State Generate explanations for each score component
\State \textbf{return} $\mathbf{s}$, explanations, source\_links
\end{algorithmic}
\end{algorithm}

\begin{figure}[!t]
\centering
\includegraphics[width=0.9\linewidth]{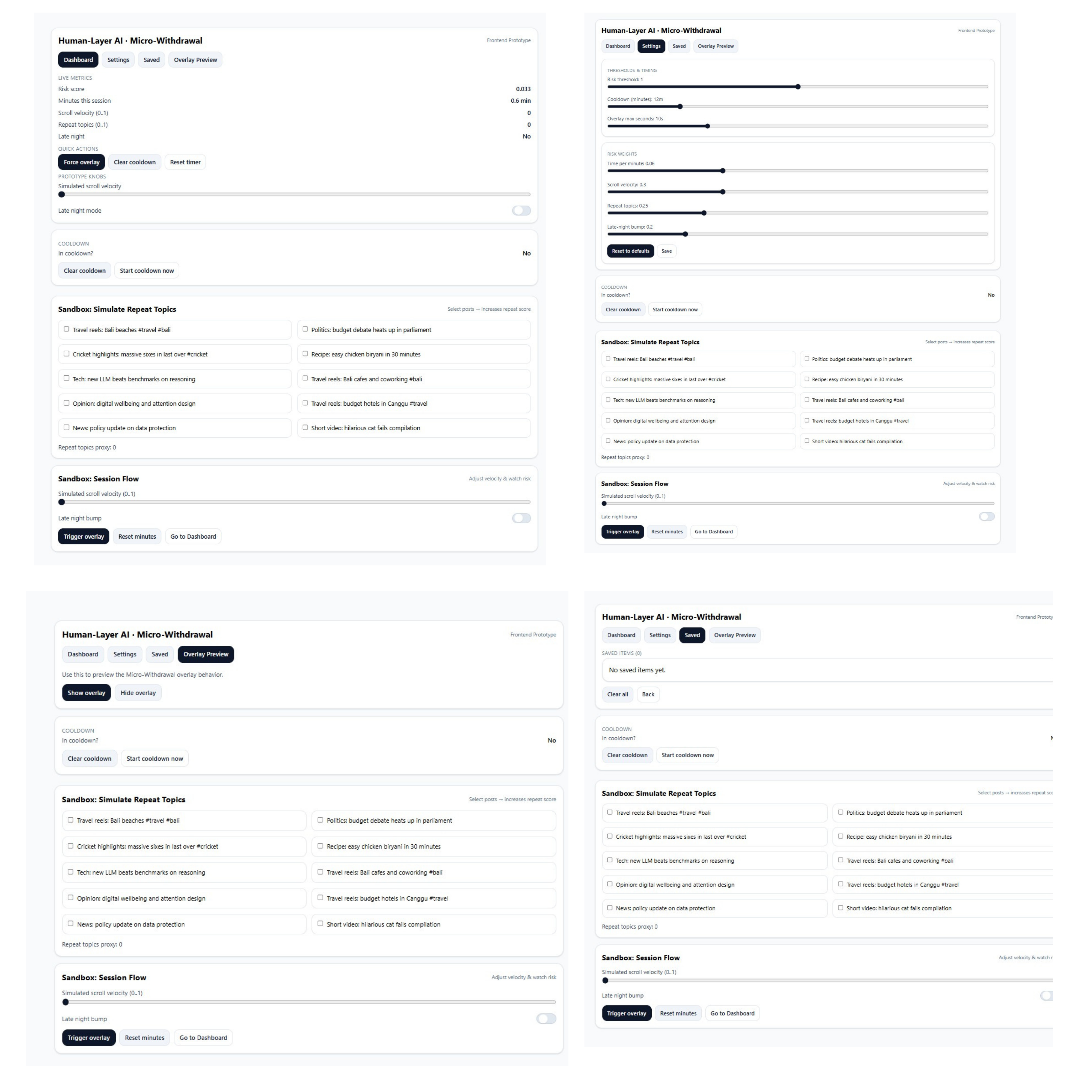}
\caption{Prototype screenshots of HL-AI browser-based intermediary (placeholder).}
\label{fig:architecture}
\end{figure}

\section{Implementation Architecture}

Our system is implemented as a Chrome/Edge browser extension that acts as a Human-Layer AI (HL-AI) intermediary, dynamically modifying the interface and injecting interaction elements without requiring platform cooperation. Figure~\ref{fig:architecture} presents the system architecture along with prototype screenshots.

The HL-AI architecture functions as a privacy-preserving intermediary that applies the design patterns without requiring any cooperation from the underlying platform. At its core, a configurable Pattern Engine implements each pattern according to user-defined parameters and thresholds, while a Context Monitor observes user state, interaction rhythms, content consumption, and temporal factors to determine when interventions are appropriate. A central Decision Coordinator resolves potential conflicts between patterns by considering safety priorities, user preferences, and situational context. The Interface Adapter then modifies how content is presented, inserts interface elements, and exposes control panels in a way that preserves the original platform’s functionality. Throughout this process, the system follows a privacy-by-design approach: personal data is processed on-device whenever possible, cloud computation is restricted to anonymized aggregates, and users maintain full control over data sharing and pattern activation.

\section{Conclusion}
We introduce a Human-Layer AI framework that reimagines social media through humane, user-controlled design patterns. Rather than optimizing for engagement, our approach centers on user sovereignty—offering protective and empowering features that foster wellbeing, safety, and agency. This pattern language outlines practical pathways for retrofitting existing platforms with ethical, user-aligned controls. By balancing autonomy with risk mitigation, it supports thoughtful decision-making that safeguards individuals while strengthening collective trust. True progress depends on embedding human-centered design principles at the core of technology. Our prototype offers an early blueprint for social media that serves human flourishing, inviting further user studies and cross-cultural evaluation to refine and expand its impact.

\bibliographystyle{IEEEtran}
\bibliography{refs}

\end{document}